\documentclass[conference]{IEEEtran}
\IEEEoverridecommandlockouts
\usepackage{cite}
\usepackage{amsmath,amssymb,amsfonts}
\usepackage{algorithmic}
\usepackage{graphicx}
\usepackage{textcomp}
\usepackage{comment}
\usepackage{xcolor}
\def\BibTeX{{\rm B\kern-.05em{\sc i\kern-.025em b}\kern-.08em
    T\kern-.1667em\lower.7ex\hbox{E}\kern-.125emX}}
\begin{document}

\author{
    \IEEEauthorblockN{
        Mohammad F. Al-Hammouri\IEEEauthorrefmark{1},
        Yazan Otoum\IEEEauthorrefmark{5},
        Rasha Atwa\IEEEauthorrefmark{3},
        Amiya Nayak\IEEEauthorrefmark{4}
    }
    \IEEEauthorblockA{\IEEEauthorrefmark{1}Dept. of Computer Engineering, The Hashemite University, Zarqa, Jordan}
    \IEEEauthorblockA{\IEEEauthorrefmark{5}School of Computer Science and Technology, Algoma University, Canada}
    \IEEEauthorblockA{\IEEEauthorrefmark{3}College of Computer Science and Engineering, University of Jeddah, Saudi Arabia}
    \IEEEauthorblockA{\IEEEauthorrefmark{4}School of Electrical Engineering and Computer Science, University of Ottawa, Canada}
}

\title{Optimizing Intrusion Detection with Hybrid Traditional and Large Language Model (LLM) Approaches\\
}

\title{Hybrid LLM-Enhanced Intrusion Detection for Zero-Day Threats in IoT Networks}

\maketitle

\begin{abstract}

This paper presents a novel approach to intrusion detection by integrating traditional signature-based methods with the contextual understanding capabilities of the GPT-2 Large Language Model (LLM). As cyber threats become increasingly sophisticated, particularly in distributed, heterogeneous, and resource-constrained environments such as those enabled by the Internet of Things (IoT), the need for dynamic and adaptive Intrusion Detection Systems (IDSs) becomes increasingly urgent. While traditional methods remain effective for detecting known threats, they often fail to recognize new and evolving attack patterns. In contrast, GPT-2 excels at processing unstructured data and identifying complex semantic relationships, making it well-suited to uncovering subtle, zero-day attack vectors. We propose a hybrid IDS framework that merges the robustness of signature-based techniques with the adaptability of GPT-2-driven semantic analysis. Experimental evaluations on a representative intrusion dataset demonstrate that our model enhances detection accuracy by 6.3\%, reduces false positives by 9.0\%, and maintains near real-time responsiveness. These results affirm the potential of language model integration to build intelligent, scalable, and resilient cybersecurity defences suited for modern connected environments.
\end{abstract}


\section{Introduction}

The exponential growth of interconnected digital systems has elevated the complexity and scale of cybersecurity threats. As organizations increasingly rely on digital infrastructure for mission-critical operations, the risk of sophisticated and undetected intrusions escalates. The increasing adoption of Internet of Things (IoT) devices in critical sectors, such as healthcare \cite{otoum2025differential}, industrial automation, and smart cities, has expanded the attack surface and complexity of securing digital infrastructures. IoT networks are characterized by their heterogeneity, constrained resources, and frequent communication over public or unsecured channels, making them particularly susceptible to zero-day exploits and multi-vector attacks \cite{otoum2024advancing}. These conditions demand IDS solutions that are lightweight, responsive, context-aware, and capable of semantic reasoning \cite{brown2020language, said2023scalable}. 
IDSs have thus become foundational to network security architectures, aiming to identify and neutralize unauthorized access, malicious activities, and data exfiltration attempts. Traditional Intrusion Detection Systems (IDSs), primarily classified into signature-based and anomaly-based systems, have provided a reliable first-line defence. Signature-based systems rely on predefined attack patterns, making them effective for known threats, whereas anomaly-based systems leverage statistical models and behavioural baselines to detect deviations. However, both approaches have limitations, particularly in adapting to zero-day exploits and polymorphic attacks that deviate from established patterns \cite{sommer2010outside}. Artificial Intelligence (AI) and Machine Learning (ML) have been extensively explored in recent years to address these challenges. Techniques ranging from decision trees to deep neural networks have been employed to learn patterns from traffic and log data, thereby improving detection performance \cite{salman2020review}. Yet, many of these models struggle with contextual understanding and generalization beyond their training distributions. In this context, LLMs, such as OpenAI's GPT-2, represent a paradigm shift. Initially developed for natural language processing tasks, LLMs have demonstrated remarkable capabilities in understanding complex semantics, identifying contextual anomalies, and generalizing across domains \cite{otoum2025llm}. This paper proposes a hybrid IDS framework that integrates the robustness of traditional detection mechanisms with the semantic intelligence of LLMs. By combining the deterministic accuracy of signature-based methods and the adaptive strengths of GPT-2, the proposed system aims to reduce false positives while maintaining high detection accuracy, especially against novel threats. This integration enables context-aware intrusion analysis, leveraging the linguistic patterns learned by LLMs to interpret network events more holistically. In contrast to prior works that treat ML and NLP models as standalone enhancements, our approach fuses conventional IDS heuristics with deep language-based analysis to build a layered, synergistic defence. This improves threat detection coverage, reduces response latency, and enhances the model's interpretability. The key contributions of this study are threefold: (i) the development of a novel hybrid intrusion detection framework that fuses traditional IDS and LLM-based contextual reasoning, (ii) a comprehensive evaluation demonstrating improved accuracy and a substantial reduction in false positives using the CSE-CIC-IDS2018 dataset, and (iii) a performance analysis that validates the system's suitability for near real-time applications through latency and AUC-ROC metrics. The remainder of this paper is structured as follows: Section II provides a comprehensive background and literature review on IDS and LLM applications in cybersecurity. Section III details the proposed hybrid methodology, including the system architecture and decision fusion strategy. Section IV describes the experimental setup and evaluates the performance using real-world datasets and the results. Finally, Section V concludes the paper and outlines directions for future research.

\section{Background}

\subsection{Literature Review}

Several researchers have proposed enhancements to traditional IDS by integrating modern intelligence-driven techniques. For example, the authors in \cite{MODI201342} explored signature-based and anomaly-based methods as foundational IDS approaches. However, their work revealed that these methods often fall short in dealing with the dynamic and rapidly evolving nature of modern cyber threats, especially those capable of bypassing static rule-based defences. The work \cite{radford2019language} introduced the GPT series of LLMs, demonstrating remarkable capabilities in understanding and generating human-like text. Their research opened the door to applying LLMs in domains such as cybersecurity, where analyzing and interpreting unstructured data is crucial. The authors in \cite{apruzzese2018effectiveness} later proposed training LLMs on network traffic and system logs to recognize patterns indicative of malicious activity. This approach effectively identified known threats and detected novel and complex attack vectors, thereby reducing false positives and enhancing IDS robustness. Furthermore, the authors of \cite{zhang2019deep} have suggested that LLMs such as GPT-2 and GPT-3 offer substantial advantages for real-time threat detection due to their ability to learn and adapt continuously. Their work emphasized that integrating LLMs into cybersecurity infrastructures can foster more proactive and intelligent defence mechanisms capable of evolving alongside threat landscapes. In the context of the IoT, the authors in \cite{yang2025hybrid} proposed a hybrid IDS tailored for resource-constrained IoT environments, combining lightweight ML models with rule-based filtering to enhance detection efficiency. Meanwhile, the paper \cite{alazab2023deep} examined various deep learning-based IDS architectures specifically for IoT networks, noting that their adaptability and scalability make them promising candidates for detecting complex multi-vector attacks in heterogeneous IoT infrastructures. At some point, the authors of \cite{sharma2024edge} presented an edge-based IDS framework that utilizes federated learning to minimize latency and safeguard data privacy across IoT nodes. These studies highlight the pressing need for adaptable and context-aware IDS solutions that address the specific constraints and attack surfaces of IoT deployments. These efforts underscore a growing consensus in the research community: that traditional IDS approaches can be significantly enhanced through the intelligent, context-aware reasoning capabilities of LLMs and the application of hybrid AI strategies tailored for modern IoT ecosystems \cite{otoum2025llms}.

\subsection{Traditional Detection}

\subsubsection{Signature-Based Detection (SBD)}

Longstanding and widely implemented techniques in IDS. This method relies on a database of predefined patterns or signatures of known threats, similar to antivirus software. Each signature contains a set of rules about a specific attack, which could include byte sequences in network traffic, known malicious instruction sequences used by malware, or characteristic headers in email spam. The operational model of signature-based IDS can be mathematically represented as a function:
\begin{equation}
\text{Detect}(x) = 
\begin{cases} 
1 & \text{if } x \in S \\
0 & \text{otherwise}
\end{cases}
\end{equation}
where $x$ represents a piece of network data, and $S$ is the set of signatures. The function returns 1 if the data matches a signature (indicating a detection), and 0 otherwise. Despite its reliability in detecting known threats, this method performs poorly against zero-day exploits or polymorphic threats that do not match existing signatures. The effectiveness of a signature-based IDS is heavily dependent on the currency and comprehensiveness of its signature database.

\subsubsection{Anomaly-Based Detection (ABD)}
Anomaly-based detection systems, by contrast, are designed to identify unusual patterns or behaviours that deviate from a norm. This approach does not rely on known signatures but instead uses machine learning algorithms to model a system's normal behaviour. Any significant deviation from this baseline behaviour is flagged as potentially malicious. The typical process involves creating a profile of regular activities over time and using statistical methods to define normal behaviour. The detection function can be outlined as follows:
\begin{equation}
\text{AnomalyScore}(x) = \text{Distance}(x, \mu)
\label{eq:anomaly-score}
\end{equation}

\begin{equation}
\text{Detect}(x) = 
\begin{cases} 
1 & \text{if } \text{AnomalyScore}(x) > \theta \\
0 & \text{otherwise}
\end{cases}
\label{eq:anomaly-detect}
\end{equation}

Here, $x$ is a new observation,  $\mu$  represents the established profile of normal behaviour, and  $\theta$  is a threshold determined during the training phase. The function Distance quantifies how far x  is from the expected behaviour $\mu$, and if this score exceeds  $\theta$, the system flags it as an anomaly. Anomaly-based systems are particularly effective against previously unknown threats and can adapt to environmental changes by updating the behavioural profile. However, they can be prone to higher false-positive rates, especially in highly dynamic networks where normal behaviour frequently changes. In practice, hybrid systems that combine signature-based and anomaly-based detection are commonly used to leverage the strengths of each approach. Such systems utilize signature-based detection for high accuracy and speed in identifying known threats and anomaly detection for their robustness against novel attacks. The integration of these methods involves an ensemble approach where outputs from both systems are combined to improve the overall detection accuracy:
\begin{equation}
\text{HybridDetect}(x) = \max(\text{DetectSBD}(x), \text{DetectABD}(x))
\label{eq:hybrid-detect}
\end{equation}

This function uses the maximum operator to combine the decisions from the signature-based and anomaly-based models, ensuring that the system flags a threat if either method detects one.

\subsection{Large Language Models}

LLMs are advanced AI systems that learn from massive text corpora to understand and generate human language. Built on Transformer architectures, they have transformed Natural Language Processing (NLP) by efficiently handling long sequences at scale. With deep networks and billions of parameters, LLMs capture grammar, context, and semantics, enabling them to perform diverse tasks, such as translation, summarization, question answering, and text generation, without requiring task-specific programming. A prominent example is OpenAI's Generative Pre-trained Transformer (GPT) series. GPT-1 (117M parameters) trained on BooksCorpus demonstrated the effectiveness of unsupervised pre-training for generalization across tasks. GPT-2 scaled this further with 1.5B parameters and training on WebText (8M pages), achieving state-of-the-art performance in generating coherent, human-like text. Its architecture is designed as follows:
\begin{itemize}
\item Model Architecture: GPT-2 utilizes a stacked Transformer model architecture, comprising multiple layers of self-attention and feed-forward neural networks. Each layer processes the input sequence and passes its output to the next layer, allowing the model to build a deep representation of the input data.
\item Training Objective: The training objective of GPT-2 is to predict the next word in a sentence given all the previous words, optimizing the likelihood of the next word using the cross-entropy loss function. This training is performed across a diverse range of internet texts, enabling the model to learn a broad array of language patterns and styles.
\end{itemize}

\section{Methodology}
The goal of GPT-2 is to train a word vector model with stronger generalization ability. It does not significantly innovate or alter the structure of GPT-1's network; instead, it utilizes more network parameters and a larger dataset. So, we will first introduce how GPT-1 works.
\subsection{GPT-1}

The training of GPT-1 is divided into unsupervised pre-training and supervised model fine-tuning. GPT-1 uses maximum likelihood estimation to train its neural networks. Given a sequence of tokens \( U = \{u_1, \dots, u_n\} \), the likelihood of the sequence is given by the following formula:
\begin{equation}
L_1(U) = \sum_i \log P(u_i \mid u_{i-k}, \dots, u_{i-1}; \Theta)
\label{eq:L1}
\end{equation}
where k  is the context window size,  P  is the probability function, and  $\Theta$  are the model parameters. The model parameters are optimized using the SGD algorithm. In GPT-1, the model consists of 12 transformer blocks. Each block is essential to the transformer architecture and contributes to the model's powerful capabilities.

\begin{figure}[ht]
\begin{center}
\advance\leftskip-3cm
\advance\rightskip-3cm
\includegraphics[keepaspectratio=true,scale=0.7]{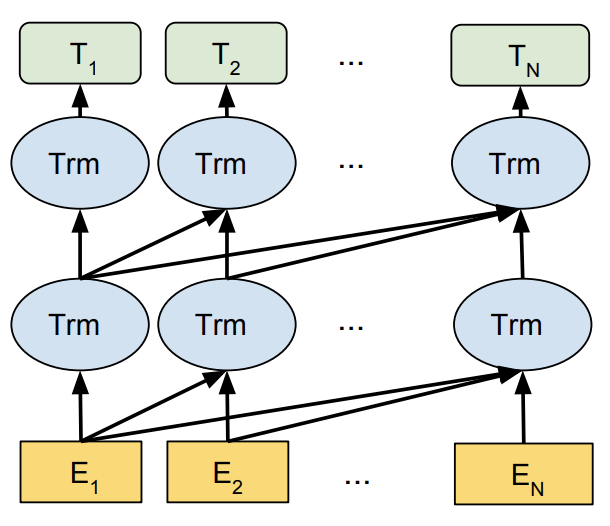}
\caption{GPT-based Next Token Language Model}
\label{fig::Figure3}
\end{center}
\end{figure}

\vspace{-1.5em}

\begin{equation}
h_0 = U W_e + W_p
\label{eq:h0}
\end{equation}
\begin{equation}
h_l = \text{transformer\_block}(h_{l-1}) \quad \forall l \in [1, n]
\label{eq:hl}
\end{equation}
\begin{equation}
P(u) = \text{softmax}(h_n W_e^T)
\label{eq:softmax}
\end{equation}

Here, \( U = (u_{1}, \dots, u_{k}) \) denotes the input sequence of tokens, \( n \) is the sequence length, \( W_e \) is the token embedding matrix, and \( W_p \) is the positional encoding matrix used to preserve token order information within the sequence. During optimization, the model computes the sequence loss over a sample \( C \), which consists of \( m \) tokens \( \{x_1, \dots, x_m\} \), where the objective is to predict the next token \( y \). The architectural structure of the GPT-based next-token prediction model, including token embedding, transformer layers, and output generation, is illustrated in Fig.~\ref{fig::Figure3}. The conditional probability of the target token \( y \) given the preceding context is defined as:

\begin{equation}
P(y \mid x_1, \dots, x_m) = \text{softmax}(h^m_l W_y)
\label{eq:conditional-softmax}
\end{equation}
where \( W_y \) is the matrix associated with the target \( y \). The following formula is used to calculate the sequence loss:
\begin{equation}
L_2(C) = \sum_{x,y} \log P(y \mid x^1, \dots, x^n)
\label{eq:loss-l2}
\end{equation}

To further optimize the loss function \( L_2 \), we incorporate it with \( L_1 \) from a different process, adding them with a coefficient \( \lambda \) to adjust their relative importance:
\begin{equation}
L_3(C) = L_2(C) + \lambda L_1(C)
\label{eq:loss-l3}
\end{equation}

Note: The softmax function uses the matrix \( W_y \) without any delimiter (delimiter) symbol.

\subsection{GPT-2}

GPT-2 is designed to perform a variety of supervised tasks by leveraging knowledge learned during unsupervised pre-training. Owing to the sequential nature of textual data, the probability of a sequence can be expressed as a product of conditional probabilities:
\begin{equation}
p(s_1, s_2, \dots, s_n) = \prod_{i=1}^n p(s_i \mid s_1, s_2, \dots, s_{i-1})
\label{eq:autoregressive}
\end{equation}

GPT-2 adopts an autoregressive framework, generating coherent outputs by conditioning on previously seen tokens. This enables it to handle diverse NLP tasks, such as question answering, summarization, and classification, by reformulating them as conditional text generation problems through the use of Prompt Engineering. The TACL Decathlon NLP (TDecNLP) competition popularized this paradigm, where models like MQAN achieved strong generalization across ten tasks using a unified architecture. GPT-2 advances this concept, demonstrating that large-scale language models trained on diverse corpora can perform new tasks without task-specific fine-tuning. For instance, when prompted with “Who is the best basketball player in history?”, the model may generate “Michael Jordan,” showcasing zero-shot inference. GPT-2’s architecture and training enable it to internalize patterns that support robust generalization, reframing supervised tasks as language modelling objectives.


\subsection{Hybrid IDS}
Integrating traditional intrusion detection methods with GPT-2 can be conceptualized as a multi-layered approach, where each technique complements the other by leveraging its respective strengths. This hybrid system aims to enhance detection accuracy, reduce false positives, and adapt dynamically to new threats. The outputs from traditional IDS (both SBD and ABD) and GPT-2 can be combined using a decision-level fusion strategy. This fusion is designed to leverage the high accuracy of traditional methods for known threats and the adaptability of GPT-2 for unknown patterns:
\begin{equation}
\text{HybridDetect}(x) = \max(\text{SBD}(x), \text{ABD}(x), \text{GPT2}(x))
\label{eq:hybrid-detect}
\end{equation}

Here, \text{GPT2}(x) is a function that assesses whether the contextual analysis by GPT-2 indicates a threat, converting its output into a binary decision. This function estimates the probability of threat-related keywords or phrases being generated in response to the input sequence:
\begin{equation}
\text{GPT2}(x) = 
\begin{cases} 
1 & \text{if } P(\text{threat} \mid x) > \tau \\
0 & \text{otherwise}
\end{cases}
\label{eq:gpt2-detect}
\end{equation}

GPT-2 was used in a zero-shot capacity without fine-tuning. Tokenized network logs were framed as input prompts, and threat probabilities were inferred based on the model's generated likelihood scores. The threshold $\tau$ is optimized over time using gradient descent based on performance metrics.

\section{Experiments}
\subsection{Experimental Design}
The CSE-CIC-IDS2018 dataset \cite{iman_sharafaldin_arash_habibi_lashkari_ali_a__ghorbani_2022} was utilized for this experiment. The Canadian Institute for Cybersecurity has developed this dataset, which features a comprehensive range of simulated attack scenarios that accurately reflect modern intrusion tactics, combined with benign traffic. The dataset was chosen due to its detailed labelling and diverse array of attack vectors, making it an excellent candidate for training and evaluating an IDS based on the GPT model. The initial step in dataset preparation involved splitting the data into two sets: 70\% for training and 30\% for testing, ensuring a substantial amount of data for both training the model and evaluating its performance under varied conditions. Before feeding the data into the model, a series of preprocessing steps was undertaken:
\begin{itemize}
\item Categorical Data Conversion: Network event features categorized as strings were converted into numerical identifiers.
\item Feature Normalization: Numerical features were scaled using Min-Max scaling to ensure that feature magnitudes did not bias the model.
\item Tokenization: Network traffic logs were transformed into tokens. These tokens represent discrete parsed elements from the logs, such as IP addresses, timestamps, and protocol types, which are crucial for understanding the context of network communications. While the use of tokenized log data is reasonable, GPT-2 is pre-trained on natural language corpora. Strengthening this aspect would benefit from clarifying whether the network logs were converted into descriptive text or structured prompts to better align with the model’s linguistic expectations.

\end{itemize}


\begin{table}[h]
\caption{Model Hyperparameters for Hybrid IDS}
\centering
\begin{tabular}{|l|l|}
\hline
\textbf{Parameter}          & \textbf{Value}                                  \\ \hline
GPT-2 Configuration         & Pre-trained, adjusted for IDS                   \\ \hline
Learning Rate               & Initial: $5 \times 10^{-5}$, dynamically adjusted \\ \hline
Batch Size                  & 32                                              \\ \hline
Epochs                      & Up to 10, with early stopping                   \\ \hline
\end{tabular}

\label{tab:model_hyperparameters}
\end{table}

\begin{figure*}[ht!]
    \centering
    \includegraphics[width=0.7\linewidth]{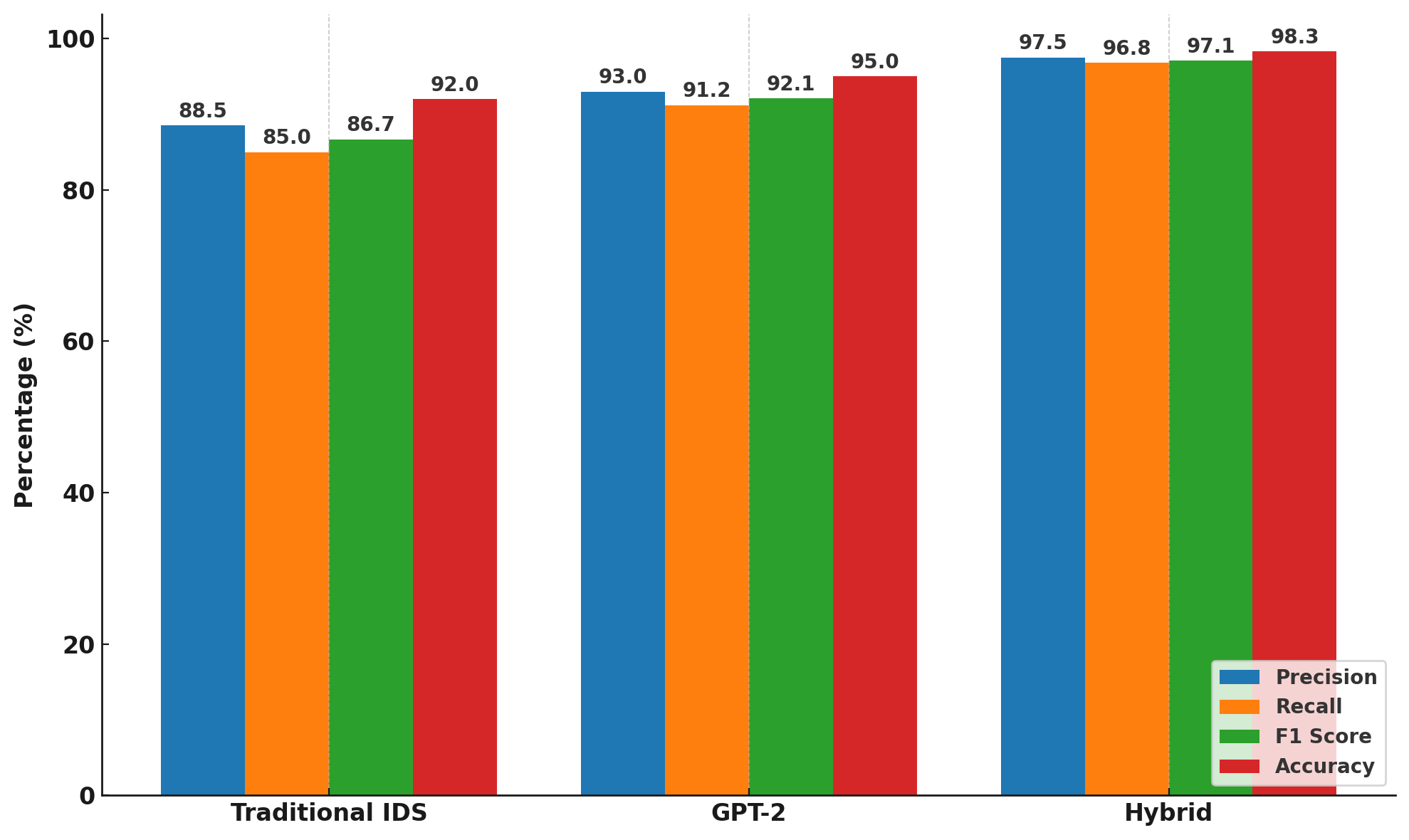}
    \caption{Performance metrics comparison across Traditional IDS, GPT-2, and Hybrid models.}
    \label{fig::results}
    \vspace{-1em}
\end{figure*}

Table~\ref{tab:model_hyperparameters} presents the hyperparameters set for the hybrid model used in our intrusion detection system experiment. These settings were carefully chosen to optimize the model’s performance for detecting network intrusions. The table details the hybrid model's configuration, learning rate adjustments, batch size, and training epochs. To thoroughly evaluate the effectiveness of our hybrid GPT-2 and traditional method-based intrusion detection system, we employed a suite of metrics designed to assess various aspects of system performance in a real-world cybersecurity environment. Accuracy measures the model’s overall correctness, precision indicates the ability to avoid false positives, and recall assesses the model’s capability to identify all genuine threats. The F1 score, which balances precision and recall, is significant in environments where false positives and negatives have substantial consequences. Additionally, the Area Under the Receiver Operating Characteristic Curve (AUC-ROC) provides a comprehensive view of the model’s ability to discriminate between classes across different thresholds, a crucial metric for binary classification tasks. The performance of our hybrid model was compared with that of traditional intrusion detection systems, which rely on signature-based or anomaly-based detection methods. This comparison is essential for demonstrating the advanced capabilities and improvements by integrating GPT-2 with traditional detection techniques. 

\subsection{Experimental Results}

The experimental evaluation of our hybrid GPT-2 and traditional methods-based intrusion detection system produced compelling results, indicating a significant enhancement in detection capabilities over traditional IDS approaches. Utilizing the CSE-CIC-IDS2018 dataset, the hybrid model demonstrated high accuracy, precision, recall, and F1 score, suggesting robust performance across all major metrics. Fig.~\ref{fig::results} illustrates a comparative analysis of the performance metrics Precision, Recall, F1 Score, and Accuracy across three intrusion detection approaches: Traditional IDS, GPT-2, and the proposed Hybrid model. As shown, the Hybrid model consistently outperforms both the traditional and LLM-only setups across all metrics. It achieves a Precision of 97.5\%, Recall of 96.8\%, F1 Score of 97.1\%, and Accuracy of 98.3\%, reflecting its enhanced ability to identify threats while minimizing false alarms accurately. The GPT-2 model also performs significantly better than the Traditional IDS, especially in Recall (91.2\% vs. 85.0\%) and F1 Score (92.1\% vs. 86.7\%), underscoring the LLM’s strength in semantic reasoning and pattern detection. However, it still falls short of the Hybrid model, indicating that the combination of rule-based reliability and contextual understanding yields a more robust and generalizable detection system. These results substantiate the effectiveness of integrating LLMs with traditional methods for building adaptive, accurate, and future-proof cybersecurity defences.

\begin{figure}[ht]
\centering
\includegraphics[width=0.9\linewidth]{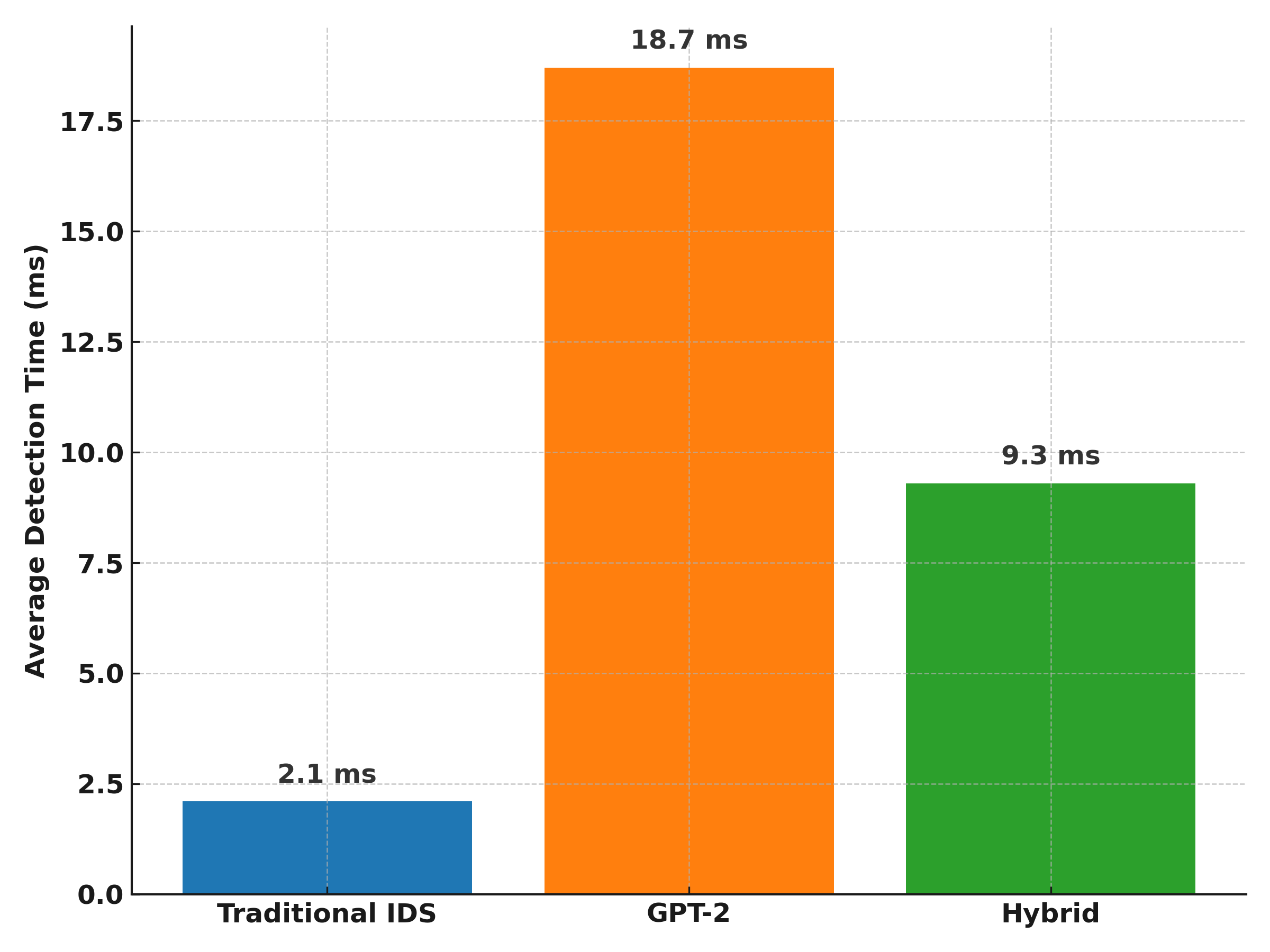}
\caption{Average detection time per sample for different IDS approaches.}
\label{fig::latency}
\end{figure}

Fig.~\ref{fig::latency} illustrates the average detection time per sample for the three evaluated intrusion detection approaches: \textit{Traditional IDS}, \textit{GPT-2}, and the proposed \textit{Hybrid} model. The Traditional IDS demonstrates the lowest latency at approximately 2.1 milliseconds, followed by the Hybrid model at 9.3 milliseconds and GPT-2 at 18.7 milliseconds. Although the Traditional IDS exhibits the lowest detection latency, this comes at the cost of limited adaptability to novel threats. The proposed Hybrid model balances latency and intelligence, offering near real-time performance (9~ms) while significantly enhancing detection accuracy. This trade-off represents a practical and scalable advancement over traditional IDS solutions, particularly for environments requiring robust defences against zero-day and evolving threats. While more computationally intensive, the GPT-2-based model delivers enhanced detection performance but with latency nearly double that of the Hybrid model. This underscores the advantage of the proposed framework, which strategically fuses traditional efficiency with LLM-driven intelligence to achieve both responsive and resilient cybersecurity.

\begin{figure}[ht]
\centering
\includegraphics[width=0.9\linewidth]{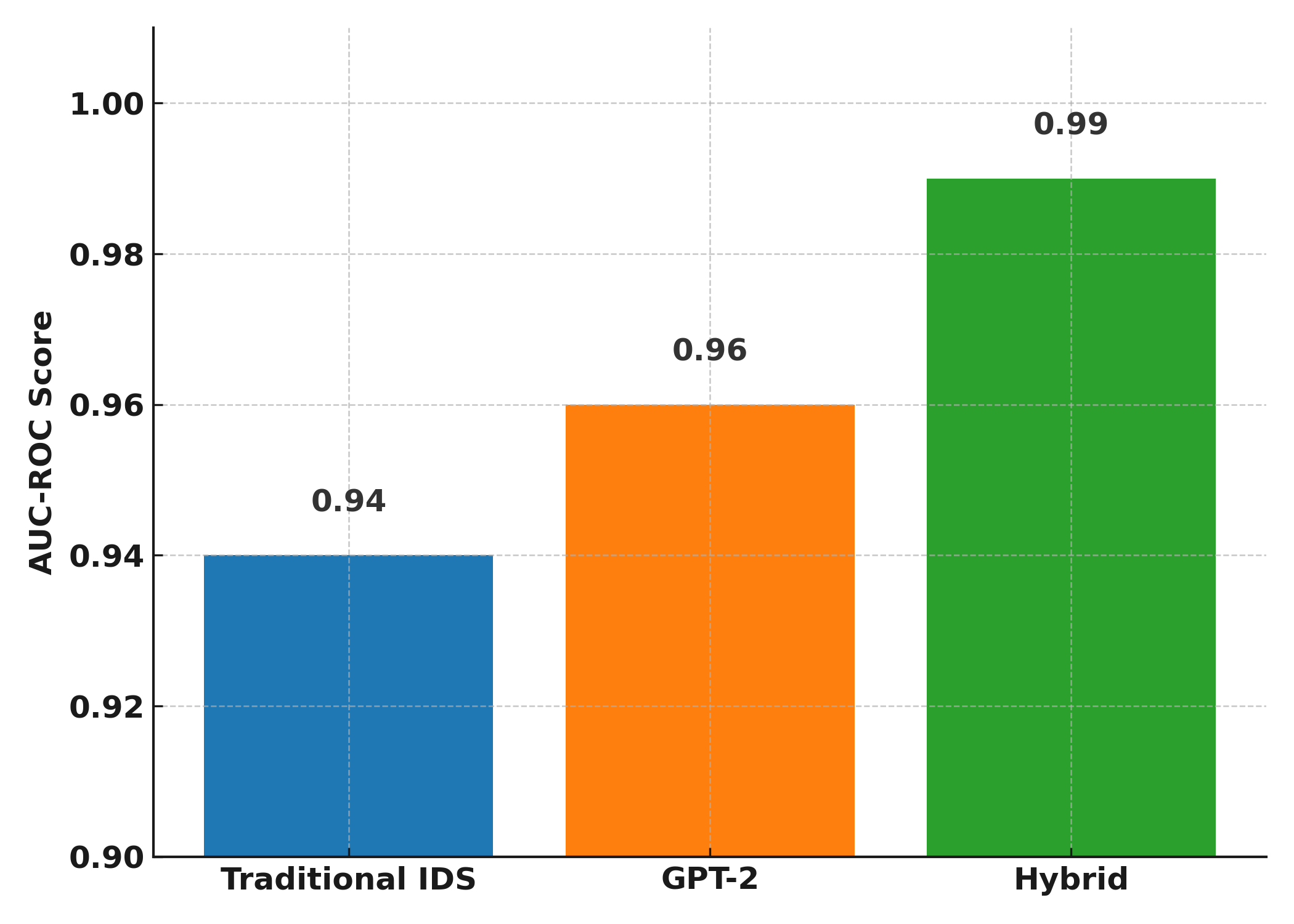}
\caption{AUC-ROC score comparison between Traditional IDS and the Hybrid model.}
\label{fig:aucroc}
\vspace{-0.5em}
\end{figure}

Fig.~\ref{fig:aucroc} compares the Area Under the Receiver Operating Characteristic (AUC-ROC) scores of the proposed Hybrid model, GPT-2-based IDS, and the baseline Traditional IDS. The AUC-ROC score is a critical metric for evaluating a model’s ability to distinguish between attack and benign traffic across varying decision thresholds. As depicted, the Hybrid model achieves an AUC-ROC of 0.99, followed by GPT-2 at 0.96 and the Traditional IDS at 0.94. This progressive improvement highlights the enhanced discriminative power of models that incorporate language model-based intelligence. The near-perfect AUC-ROC of the Hybrid model demonstrates its effectiveness in both identifying true intrusions and minimizing false positives, even under varying classification conditions. These results validate the robustness and adaptability of the proposed framework, emphasizing its suitability for deployment in complex and dynamic cybersecurity environments. The high AUC-ROC of 0.99 further confirms the model’s robustness across decision thresholds, indicating its generalizability across different intrusion scenarios with minimal compromise on either false positives or false negatives. Collectively, these results demonstrate that the proposed hybrid intrusion detection framework significantly outperforms conventional methods in both accuracy and adaptability. By fusing the structured detection strength of traditional IDS with the semantic intelligence of GPT-2, the model establishes a scalable and future-ready approach to proactive cyber defence.

\section{Conclusion}

This paper presented a hybrid intrusion detection framework that integrates traditional signature-based methods with the semantic intelligence of GPT-2 to enhance cybersecurity resilience, particularly in IoT-driven environments. Through extensive experiments on the CSE-CIC-IDS2018 dataset, the proposed model demonstrated substantial improvements over conventional IDS approaches, achieving 98.3\% accuracy, a 0.99 AUC-ROC score, and marked reductions in false positives. These results validate the effectiveness of combining rule-based precision with LLM-driven contextual analysis in detecting both known and novel threats more efficiently. The hybrid design not only preserves near real-time responsiveness but also enhances detection adaptability, making it especially suitable for dynamic, resource-constrained infrastructures such as IoT networks. 
Looking ahead, we plan to refine the proposed IDS framework to address the inherent limitations of IoT environments better. Specifically, future work will focus on improving energy efficiency and computational scalability to support deployment on lightweight edge devices. These enhancements will help ensure the model’s viability for large-scale, real-time intrusion detection across heterogeneous IoT networks, where responsiveness and minimal resource consumption are essential.


\end{document}